\documentclass[sn-mathphys-num]{sn-jnl} 

\newcommand{\ptx}{\ensuremath{p_{T}}\xspace}
\newcommand{\pta}{\ensuremath{p_{T1}}\xspace}
\newcommand{\ptb}{\ensuremath{p_{T2}}\xspace}
\newcommand{\sqrtsNN}{\ensuremath{\sqrt{s_{{\rm NN}}}}\xspace}

\newcommand{\NPartAv}{${\rm \langle N_{Part} \rangle}$}
\newcommand{\RAv}{$R^{\rm eff}$}
\newcommand{\LAAv}{$\langle L_{1} \rangle$}
\newcommand{\LBAv}{$\langle L_{2} \rangle$}
\newcommand{\SNNA}{(I)}
\newcommand{\SNNB}{(II)}

\newcommand{\SNNO}{$\sqrt{s_{NN}}$ = 2.76 TeV}
\newcommand{\SNNT}{$\sqrt{s_{NN}}$ = 5.02 TeV}
\newcommand{\npart}{N_{\rm Part}}

\usepackage{verbatim}
\usepackage{graphicx}
\usepackage{rotating}
\usepackage{graphicx}% Include  files
\usepackage{dcolumn}% Align table columns on decimal point
\usepackage{bm}% bold math
\usepackage{epsfig}
\usepackage{hyperref}
\usepackage{ulem}
\usepackage{appendix}
\usepackage{mwe}

\expandafter\let\csname equation*\endcsname\relax
\expandafter\let\csname endequation*\endcsname\relax
\usepackage{amsmath}
\usepackage{amssymb}
\usepackage[caption=false]{subfig}

\begin{document}
\title[DiJet]{Quantifying the Jet Energy Loss in Pb+Pb collisions at LHC} 
%\author*[1,2]{\fnm{Vineet} \sur{Kumar}}\email{vineetk@barc.gov.in}
\author*[1,2]{\fnm{Vineet} \sur{Kumar}}
\author*[1,2]{\fnm{Prashant} \sur{Shukla}}\email{pshukla@barc.gov.in}
\affil*[1]{\orgdiv{Nuclear Physics Division}, \orgname{Bhabha Atomic Research Center},
  \orgaddress{\city{Mumbai}, \postcode{400085}, \state{Maharashtra}, \country{India}}}

\affil[2]{\orgname{Homi Bhabha National Institute},
  \orgaddress{\street{Training School Complex}, \city{Anushaktinagar, Mumbai},
    \postcode{400094}, \state{Maharashtra}, \country{India}}}

\date{\today}

\abstract{
  In this work, we give a method to study the energy loss of jets in the medium
  using a variety of jet energy loss observables such as nuclear modification factor
  and transverse momentum asymmetry in dijets and $\gamma$-jets in heavy ion collisions. 
  The energy loss of jets in the medium depends mainly on the size and the properties
  of medium viz. temperature and is a function of energy of the
  jets as predicted by various models. A Monte Carlo (MC) method is employed
  to generate the transverse momentum and path-lengths of the initial jets
  that undergo energy loss.
  %A Monte Carlo method is used to generate transverse momentum and path-lengths
  %of the initial jets which suffer energy loss.
  %Simulations are conducted to compare the measurements of various factors, such as nuclear modification factors,
  %dijet momentum imbalance measures, and gamma-jet asymmetry, at energy levels of 2.76 TeV and 5.02 TeV. These
  %comparisons involve the utilization of different energy loss scalings, as well as considering the dependencies
  %on transverse momentum and system size.
  Using different scenarios of energy loss, the transverse momentum and system size
  dependence of nuclear modification factors and different measures of dijet momentum imbalance
  at energies $\sqrt{s_{\rm NN}}$ = 2.76 TeV and 5.02 TeV and $\gamma$-jet asymmetry
  at $\sqrt{s_{\rm NN}}$ =2.76 TeV in Pb+Pb collisions are simulated. The results
  are compared with the measurements by ATLAS and CMS experiments as a function
  of transverse momentum and centrality.
  The study demonstrates how the system size and energy
  dependence of jet energy loss can be quantified using various experimental
  observables.}

%PACS codes here, in the form: \PACS code \sep code
%

%\begin{keyword}
%  PbPb \sep Quark-gluon plasma \sep Dijet \sep LHC
%  \PACS 12.38.Mh \sep 24.85.+p
%\end{keyword}

%
%Uncomment for keywords JPG
%\noindent{\it Keywords}: quark-gluon plasma, LHC, Jet


%\pacs{12.38.Mh, 24.85.+p}
%\keywords{quark-gluon plasma, LHC, Jet}
%

%\submitto{\JPG}

%\maketitle
% 
% For two-column output uncomment the next line and choose [10pt] rather than [12pt]
%in the \documentclass declaration

%\end{frontmatter}

\maketitle

\newpage

\section{Introduction}
\label{Sec:Introduction}

% Normal QGP introduction

The PbPb collisions at Large Hadron Collider (LHC) are performed to produce and
study the properties of bulk strongly interacting matter at high temperatures
where the quarks and gluons provide the basic degrees of freedom \cite{Busza_2018}.  
The matter in this phase is called quark gluon plasma (QGP) which is short-lived
but leaves its imprints on many observables which are
captured by gigantic multi particle detectors. The experimental probes
are categorised in terms of soft probes which determine the global properties of
the system and hard probes which give tomography of the system; jets are in the later
category. 
When two hadrons collide at high energies, the hard scattering of partons produces
two virtual back-to-back partons. These partons subsequently evolve as parton showers,
hadronize and are observed as two back-to-back hadronic jets.
In heavy ion collisions, the quarks and gluons produced in the hard scattering
interact strongly with the hot QCD medium due to their color charges, and lose energy,
either through the collisions with medium partons or through gluon
bremsstrahlung~\cite{Baier:1994bd, Gyulassy:1990ye}.
The jet properties are thus modified in heavy ion collisions, a phenomenon which is
termed as jet-quenching~\cite{Bjorken:1982tu}.
Since the hard partonic scattering occurs early in the collisions, the produced jets
probe the medium properties such as stopping power and transport coefficients.
  The jet-quenching can be quantified in many ways. The most conventional way
is the nuclear modification factor $R_{AA}$ which is defined as the
ratio of number of jets in heavy ion collisions to the suitably scaled number
of jets in pp collisions. The nuclear modification factor for 
light hadrons at high $p_T$ is also a measure of
jet-quenching at RHIC~\cite{STAR:2003fka} and at
LHC~\cite{ALICE:2010yje,CMS:2012aa,CMS:2016xef}.
The $R_{AA}$ values for hadrons at RHIC and the LHC are very similar although
one would expect the energy loss to increase with increased collision energy.
QCD-motivated models are generally able to describe inclusive single particle
$R_{AA}$ qualitatively.
%However, for each model the details of the
%calculations make it difficult to directly compare results between models
%and extract quantitative information about the properties of the medium
%as argued in Ref.~\cite{PHENIX:2008ove}.

%measurements of fully reconstructed jets 
The fully reconstructed jets have allowed to measure the momentum imbalance between the
leading and sub-leading jet in each event which are quantified using dijet asymmetry variables. 
Due to kinematic and detector effects, the observed energy of dijets is not
perfectly balanced, even in pp collisions. Thus to obtain a quantitative measure
of the dijet imbalance, systematic comparison of results from
heavy ion collisions with pp collisions are done.
The dijet measurements from the ATLAS~\cite{ATLAS:2010isq}
and CMS~\cite{CMS:2011iwn,CMS:2012ulu} show that, the distributions of dijet asymmetry
for peripheral PbPb collisions
are similar to those from pp collisions but are broader for central PbPb collisions.
Recently, ATLAS measured jet imbalance after background subtraction and applied
unfolding procedure to account for experimental effects~\cite{ATLAS:2017xfa,ATLAS:2022zbu}
with similar conclusions.
A recent measurement from ATLAS gives substructure-dependent jet suppression
in Pb+Pb collisions at 5.02 TeV~\cite{ATLAS:2022vii}.
%For jets at very large energies, the distribution of asymmetry tends to look similar
%to that observed in pp.
There have also been results from RHIC; The measurements
from STAR experiment~\cite{STAR:2016dfv} using a high momentum component selection
($p_{T} \geq$  2 GeV/c) observed the similar energy imbalance seen by ATLAS and CMS.
With the large statistics data collected during the PbPb runs of the LHC at 5.02 TeV,
the measurements of $\gamma$-jet and $Z^0$-jet pairs have become accessible. 
Both the photon and Z$^{0}$ do not participate in strong interactions and hence
escape the medium unattenuated and thus the energy of the paired jet can be determined
precisely.
The CMS measurement~\cite{CMS:2012ytf} of isolated photons with $p_{T}\geq$  60 GeV/c
and associated jets with $p_{T}\geq$ 30 GeV/c in
PbPb collisions at $\sqrt{s_{\rm NN}}=$ 2.76 TeV demonstrates that jet loses energy 
increasingly with increasing centrality.
Recent measurements of Z boson-tagged jets in PbPb collisions at
$\sqrt{s_{\rm NN}}=$ 5.02 TeV~\cite{CMS:2017eqd}, although with limited statistics,
show that the transverse momentum of the jet shifts to lower values.

\begin{figure*}
  \includegraphics[width=0.99\textwidth]{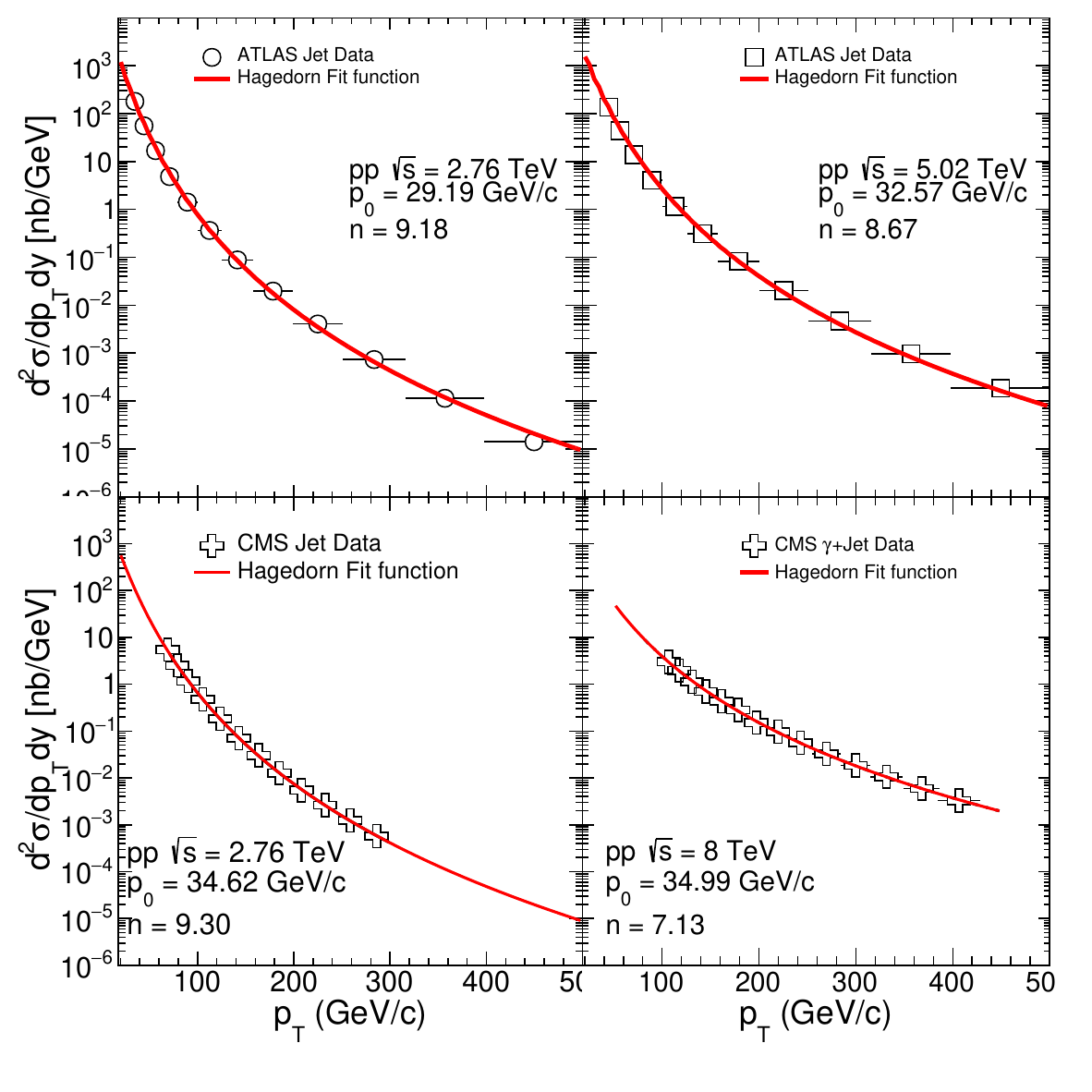}
  \caption{(Color online) Jet yields as a function of jet \ptx in pp collisions measured by
    ATLAS~\cite{ATLAS:2014ipv,ATLAS:2018gwx} and CMS experiments~\cite{CMS:2016uxf,CMS:2017eqd}. The lines correspond
    to Hagedorn fit, the parameters of which are given in the figure.}
  \label{Fig:JetPtYield}
\end{figure*}

%explain eloss formalism here
There are many approaches to describe the jet-medium interactions.
In Gyulassy-Levai-Vitev (GLV) formalism~\cite{Gyulassy:2000fs,Gyulassy:2000er},
a systematic expansion in opacity is used to extract the radiative energy loss
of partons inside QGP which has logarithmic dependence on jet
energy. 
In the Baier-Dokshitzer-Peigne-Schiff (BDPS) approach, a fast parton radiates
gluon by multiple coherent scatterings in the thick medium~\cite{Baier:1994bd}.
The finite medium effect was discussed in BDMPS work 
~\cite{Baier:1996kr,Baier:1996sk}.
Phenomenological models of parton energy loss~\cite{Muller:2002fa}
in QGP tend to define simple
dependence of the radiative energy loss of the parton on the energy of the parton
inside the medium.
The energy loss can be characterized in terms of coherence length l$_{\rm coh}$,
which is associated with the formation time of gluon radiation by a group of
scattering centres, the mean free path ($\lambda$) of the parton and the
medium size $L$ \cite{Peign:2009, Baier:2000}.
 If $l_{\rm coh} < \lambda$, the energy loss per unit length is proportional
to the energy of the parton which is known as Bethe-Heitler regime.
The regime $l_{\rm coh} > \lambda$ corresponds to  Landau-Pomeranchuk-Migdal (LPM) regime
where the energy loss per unit length is proportional to the square-root of the incoming
parton energy. If $L >> l_{\rm coh}$, the energy loss per unit length
is proportional to the square root of the energy of the parton (BDPS result).
For $ L << l_{\rm coh} $, the energy loss per unit
length is independent of energy but proportional to the parton path length.
These simple dependencies have been used to explain the charged particle spectra at RHIC
and LHC in heavy ion collisions~\cite{De:2011aa,De:2011fe}.

{\color{black}
Recently the fragmentation of partons propagating in a dense quark gluon plasma is studied using
leading double logarithmic approximation in perturbative QCD~\cite{Caucal:2018dla}. The effects of
the medium on multiple vacuum like emissions is found to modify the in medium parton showers in comparison to the
vacuum showers~\cite{Caucal:2018dla}. These modified parton showers are then used to study the jet nuclear modification
factor $R_{AA}$ and the distributions of jet splitting function Z$_{g}$ at LHC energies using Monte Carlo
methods~\cite{Caucal:2019uvr}. It is found that the energy loss by the jet is increasing with the jet transverse
momentum, due to a rise in the number of partonic sources via vacuum like emissions~\cite{Caucal:2019uvr}.
Another consequence of the jet substructure modification inside QGP will demonstrate itself through the cone size
dependence of jet energy loss. An analytical description has been obtained for the cone size dependent jet spectrum in
heavy ion collisions at the LHC energies implemented in a event by event setup including
hydrodynamic expansion of the quark gluon plasma and accounting for multiple scattering effects~\cite{Mehtar-Tani:2021fud}.
The calculation yields a good description of the centrality and p$_{T}$ dependence of jet suppression for R =
0.4 together with a mild cone size dependence, which is in agreement with recent experimental results~\cite{Mehtar-Tani:2021fud}.
}

The measured $R_{AA}$ of jets is used to obtain energy loss of jets in many 
phenomenological studies~\cite{Spousta:2016agr,Ringer:2019rfk,Ortiz:2017cul}.
Such studies assume that the energy loss is given by a power law in terms of $p_{T}$
and the value of power index is obtained by fitting the $R_{AA}$
as a function of $p_{T}$ and centrality.
In Ref.\cite{Saraswat:2017kpg}, the power law function is applied to
describe the transverse momentum distributions of charged particles and jets in
heavy ion collisions which includes the transverse flow in low $p_{\rm{T}}$ region and
the in-medium energy loss (also in terms of power law) in high $p_{\rm{T}}$ region.
The jet-quenching is quantified using jet nuclear modification factors in
various kinematic regions in terms of parameters of a power law function
assumed for energy loss~\cite{Shukla:2020mmk}.
%It should be remembered that the fragmentation changes the momentum between
%the partonic stage (at which energy is lost) and hadron formation.
%There are models which take the energy loss in account 
%at fragmentation stage~\cite{Beraudo:2012bq}.
Although there are simulation packages like JETSCAPE~\cite{JETSCAPE:2022jer} which deal with
the system more elaborately but simple models provide easy alternatives to characterize
the essential features of energy loss using different variables in different collisions systems.

\begin{figure*}
  \includegraphics[width=0.99\textwidth]{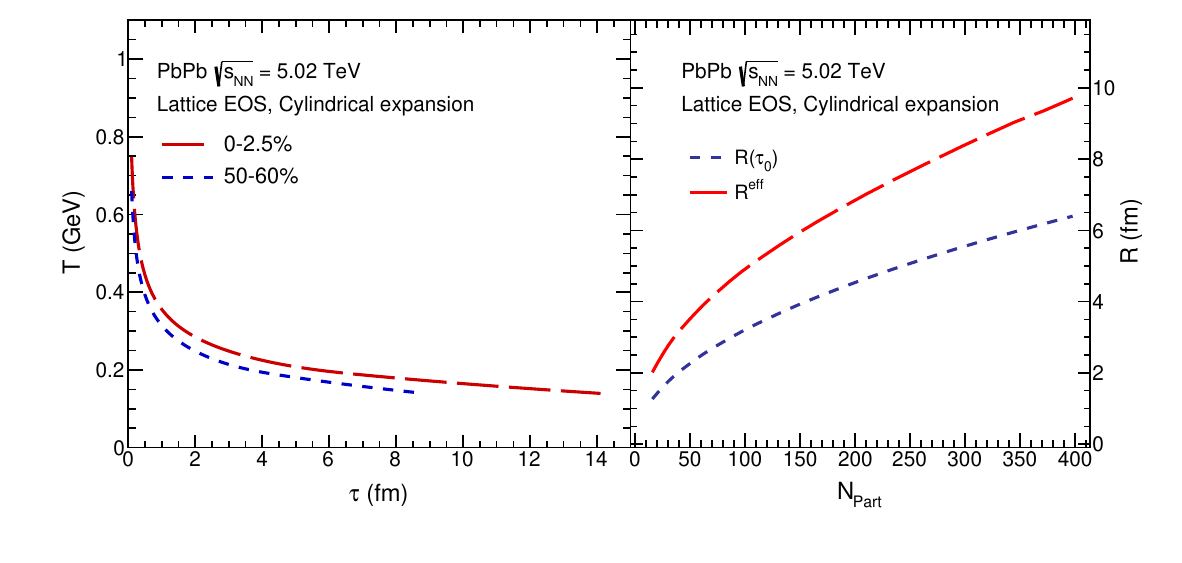}
  \caption{(Color online) (left) Temperature in the system as a function of proper time $\tau$
    in cases of the central (0-2.5$\%$) and peripheral (50-60$\%$) collisions for cylindrical expansions
    using the lattice equation of state. (right) Transverse size of the system as a function of
    $N_{\rm Part}$ of the collisions. The dashed blue line shows the transverse size, $R(\tau_0)$, at the
    initial time $\tau_{0}$ while long dashed red line shows the effective transverse size, $R^{\rm eff}$, 
    obtained by integrating the $R(\tau)$ over the time evolution of the system.}
  \label{Fig:LatticeEOS}
\end{figure*}

In this work, we have proposed a method which calculates experimental energy loss probes  namely dijet asymmetry as well as
$R_{AA}$ taking different energy loss prescriptions as inputs. The aim is to obtain the energy loss dependence on variables like
transverse momentum and centrality. We obtain the parameters of these energy loss formulas from experiments hence
quantifying the energy loss.  Once the parameters are obtained by fitting over large datasets it has excellent predictive
power in other kinematics regions. The aim of comparing the measurements at different collision energy and by different experimental
group is to bring out the underlying difference due to physics or due to different method of analysis affecting the energy loss probes.
%The study sheds light on the effectiveness of these experimental probes. 
%We did not restrict the value of alpha to below one 
%We assume that the energy loss of jets in the hot medium
%is a function of jet energy in the form of power law where the power
%index ranges from 0 (constant) to 1 (linear).
We assume that the energy loss of jets in the hot medium
is a function of jet energy in the form of power law where the power
index is extracted using the $R_{AA}$ and dijet asymmetry data from
the LHC.
%It has to be noted that the value of power law index is 1/2 in the
%BDPS approach~\cite{Baier:1994bd}.
In addition, we use (GLV) form
of energy loss which is logarithmic in energy~\cite{Gyulassy:2000fs,Gyulassy:2000er}. 
Determining the pathlenth of
the jets in the medium is equally important to know the energy loss suffered
by jets. Guided by these considerations, Monte Carlo method is employed to
generate the jet $p_T$ which travel certain pathlenth in the medium and lose energy.
Different forms of energy loss are used to obtain dijet and $\gamma$-jet asymmetry
and nuclear modification factor in the PbPb collisions at two LHC energies.

\section{Jet energy loss}
\label{Sec:JetEnergyLoss}
To describe the transverse momentum distribution of jets in pp collisions,
Hagedorn function is used which is given by ~\cite{Hagedorn:1983wk,Shukla:2020mmk}
\begin{equation}
  \frac{d^2\sigma}{dp_Tdy} = 2\pi p_T \, \frac{dn}{dy} \left(1+\frac{\ptx}{p_0}\right)^{-n}.
\label{EQhage}
\end{equation}
Figure~\ref{Fig:JetPtYield} shows the jet yields as a function of jet \ptx
in pp collisions measured by ATLAS~\cite{ATLAS:2014ipv,ATLAS:2018gwx} and CMS
experiments~\cite{CMS:2016uxf,CMS:2017eqd}. The lines correspond to Hagedorn fit. The
parameters $n$ and $p_{0}$ obtained by fitting the \ptx distribution of jets are given
on the Fig.~\ref{Fig:JetPtYield}. %The initiat jet transverse momenta (\ptx) 
%are generated as per the distribution given by Eq.~\ref{EQhage}.
{\color{black} These measured \ptx distributions parameterized with the Hagedorn fit~(Eq.~\ref{EQhage})
  along with the measured asymmetry in pp data are used to calculate the initial jets
  transverse momenta (\ptx). %the imbalance in pT is also obtained using the measured pp data.
%The XJ pp baseline is fitted to the data and used for the heavy ion case computation.
}

%The parameters $n=9.18$ and $p_{0}=29.19$ GeV/$c$ are obtained by fitting the
%\ptx distribution of jets meqasured by ATLAS~\cite{ATLAS:2014ipv}.

\begin{figure*}
  \centering
  \includegraphics[width=0.35\textwidth]{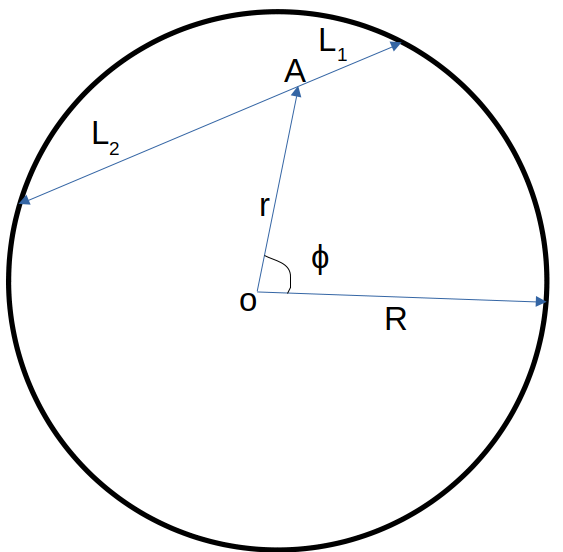}
  \caption{(Color online) Snapshot of transverse cross section of the expanding QGP medium,
    two back-to-back jets produced at radial distance $r$ traversing different
    path-lengths in the medium.}
  \label{Fig:DiJetDiagram}
\end{figure*}

The specific energy loss, $dE/dx$ is modeled as a power law
in \ptx of jet \cite{Shukla:2020mmk}
\begin{equation}
  %\frac{dE}{dx} = \frac{M}{\sqrt{p_{\rm T0}}} \left(\frac{\ptx}{p_{\rm T0}} - C\right)^{\alpha}.
  \frac{dE}{dx} = M \left(\frac{\ptx}{p_{\rm T0}} - C\right)^{\alpha}.
  \label{eqnPL}
\end{equation}
The index $\alpha$ decides the energy dependence of jet energy loss and
the parameter $M$ is dependent on medium properties such as the temperature of QGP
e.g $ \frac{M}{\sqrt{p_{\rm T0}}} \sim C_{R} \,\alpha_{S} \, \sqrt{\mu^{2} / \lambda}$ in BDPS
formalism~\cite{Baier:1994bd}. 
Here, $p_{\rm T0}$ is a scale set to 1 GeV/$c$ and value of $C=0$.

We also used GLV formalism~\cite{Gyulassy:2000fs,Gyulassy:2000er}
of parton energy loss inside QGP. 
 In this formalism, the energy loss is given by 
\begin{equation}
 \frac{\Delta E}{L} = \frac{C_{R}\,\alpha_{S}}{N(E)} \, \frac{L\mu^{2}}{\lambda} \, \log\frac{E}{\mu}.
\end{equation}
%Here, $C_{R}\,( = (N_{C}^{2}-1)/2N_{C} = 4/3) $ is color factor,
    {\color{black} Here, $C_{R}$ is the color factor for quarks and gluons.
      The QCD values of color factors are 4/3 ($C_{Rq}$) and 3 ($C_{Rg}$)  for quarks and gluons
      respectively.
      %and one naively expects that the ratio of energy loss rate
      %for quarks and gluons will be around 0.44 due to Casimir scaling~\cite{Casalderrey-Solana:2007knd} although
      %the experimental verification of this effect is inconclusive. The behavior of quark and gluon jets in the medium
      %is found to be similar and the relative energy loss of quark and gluon jets approaches a constant value which
      %is closer to unity~\cite{Apolinario:2018rhj}.
      We use a weighted average for $C_{R}$ such as 
      \begin{equation}
        %CR = (FqValPar*CR_Q) + ((1.0-FqValPar)*CR_G);
        C_{R} = f_{q} C_{Rq} + (1-f_{q})C_{Rg}
      \end{equation}
      where $f_{q}$ is \ptx dependent quark fraction calculated using PYTHIA~\cite{Sjostrand:2014zea} in ref.~\cite{Spousta:2015fca}.
    }
    The $\alpha_{S}\,(= 0.3)$ represents the QCD coupling constant, $\lambda$ is mean free path
    of parton inside the QGP, $\mu^2\,(= 4\pi\alpha_{S}T^{2}(1+N_f/6))$ is thermal gluon mass. 
    {\color{black}The factor N(E) encoded the effect of finite kinematics on radiative energy loss which
      causes it to deviate considerably from the asymptotic value of 4~\cite{Gyulassy:2000fs}.}
    The value of $N(E)$ has slow dependence at parton energy as $N(E)$ = 24.4, 10.1, 7.3, 4.0
    for $E$ = 5, 50, 500, 1000 GeV, respectively. We use linear interpolation to determine the
    values of $N(E)$ at intermediate energies.

    The mean free path $\lambda$ is given by
    \begin{equation}
\lambda^{-1} = \rho_g \sigma_{Qg} + \rho_q \sigma_{Qq},
\end{equation}
where 
\begin{eqnarray}
  \rho_{g} &=& 16~T^{3}~\frac{1.202}{\pi^{2}},~~~\rho_{q}=9~N_{f}~T^{3}~\frac{1.202}{\pi^{2}}, \\
  \sigma_{Qq} &=& \frac{9\pi\alpha_{s}^2}{2 \mu^2} \mbox{ and } \sigma_{Qg} = \frac{4}{9} \sigma_{Qq}.
\end{eqnarray}

{\color{black}The $\rho_{g}$ and $\rho_{q}$ are the quark and gluon densities inside the QCD medium, $\sigma_{Qg}$
and $\sigma_{Qq}$ represents cross section of quark quark and quark gluon interaction.}
The values of $N_{f}\,(= 3)$ and $N_{C}\,(= 3)$ are used in the
calculations~\cite{Kumar_2019}.
%At $T_{\rm eff}= 300$ MeV, the values of the parameters are
%$\mu=0.711$ GeV and $\lambda=0.86$ fm.
At $T_{\rm eff}= 0.212(0.220)$ GeV, the values of the parameters are
$\mu=0.504 (0.523)$ GeV and $\lambda=1.21 (1.17)$ fm.

The evolution of the system for each centrality class is governed by an isentropic
cylindrical expansion ($s(T)\,V(\tau)= s(T_0)\,V(\tau_0)$) with prescription given
in Ref.~\cite{Kumar:2014kfa}.
The equation of state (EOS) obtained by Lattice QCD and hadronic resonances is
used~\cite{Huovinen:2009yb}. The transverse size $R$ for a given centrality
with number of participant $\npart$ is obtained as $R(\npart,\tau_0) = R_{A} \sqrt{\npart/2A}$,
where $R_{A}$ is radius of the nucleus.
The initial entropy density, $s(\tau_0)|_{0-2.5\%}$, for 0-2.5\% centrality is 
\begin{eqnarray}
s(\tau_0)|_{0-2.5\%}  = {a_{m} \over V(\tau_0)|_{0-2.5\%}}   \left(\frac{dN}{d\eta}\right)_{0-2.5\%} . 
\label{TempVsMult}
\end{eqnarray}  
Here $a_m$ (= 5) is a constant which relates the total entropy to the total 
multiplicity $dN/d\eta$ obtained from hydrodynamic calculations~\cite{Shuryak:1992wc}.
The volume element $V(\tau)=\tau\,\pi\,(R(\tau))^{2}$ is given by
\begin{equation}
V(\tau) = \tau\,\pi\,\left( R(\tau_0) + {1\over 2} a_{T} \, \tau^2 \right)^{2},
\end{equation}
where $a_{T} = 0.1\,c^2$ fm$^{-1}$ is the transverse acceleration \cite{Kumar:2014kfa}.
We estimate the initial temperature, $T_0$, in the 0-2.5$\%$ most central collisions
from the total multiplicity in the rapidity region of interest, assuming that the initial time is
$\tau_0 = 0.1$ fm/$c$ over all rapidities. The total multiplicity in a given rapidity region is
1.5 times the charged particle multiplicity in PbPb collisions.  With the lattice
EOS, at midrapidity, with $(dN_{\rm ch}/d\eta)_{0-2.5\%} = 1943$~\cite{ALICE:2015juo}, 
we find $T_0 = 0.606$ GeV at {\sqrtsNN} = 5.02 TeV. At {\sqrtsNN} = 2.76 TeV
$(dN_{\rm ch}/d\eta)_{0-2.5\%}$ and $T_0$ are 1620 and 0.571 GeV respectively.

Figure~\ref{Fig:LatticeEOS} (left) shows temperature in the system as a function of proper time $\tau$
in case of the central (0-2.5$\%$) and peripheral (50-60$\%$) collisions for cylindrical expansions
using the lattice equation of state.
%The temperature is integrated over the time evolution of
%the system for each centrality class to extract the effective temperature, $T_{\rm eff}$. The
%$T_{\rm eff}$ is similar for all the centrality classes because longer life time of QGP
%compensate for higher initial temperature $T_{0}$ in central collisions.
The temperature is averaged over the time evolution of
the system for each centrality class to extract the effective temperature, $T_{\rm eff}$.
The $T_{\rm eff}$ is similar for all the centrality classes because longer life time of QGP
compensates for higher initial temperature $T_{0}$ in central collisions as shown
in Figure~\ref{Fig:LatticeEOS}.
The calculated values of $T_{\rm eff}$ are 0.220 GeV and 0.212 GeV at {\sqrtsNN} = 5.02
and 2.76 TeV, respectively. Our estimated values are in agreement with the 
$T_{\rm eff}$ values extracted using the hydrodynamic modelling in ref.~\cite{Gardim:2019xjs}.

\begin{figure*}
  \includegraphics[width=0.99\textwidth]{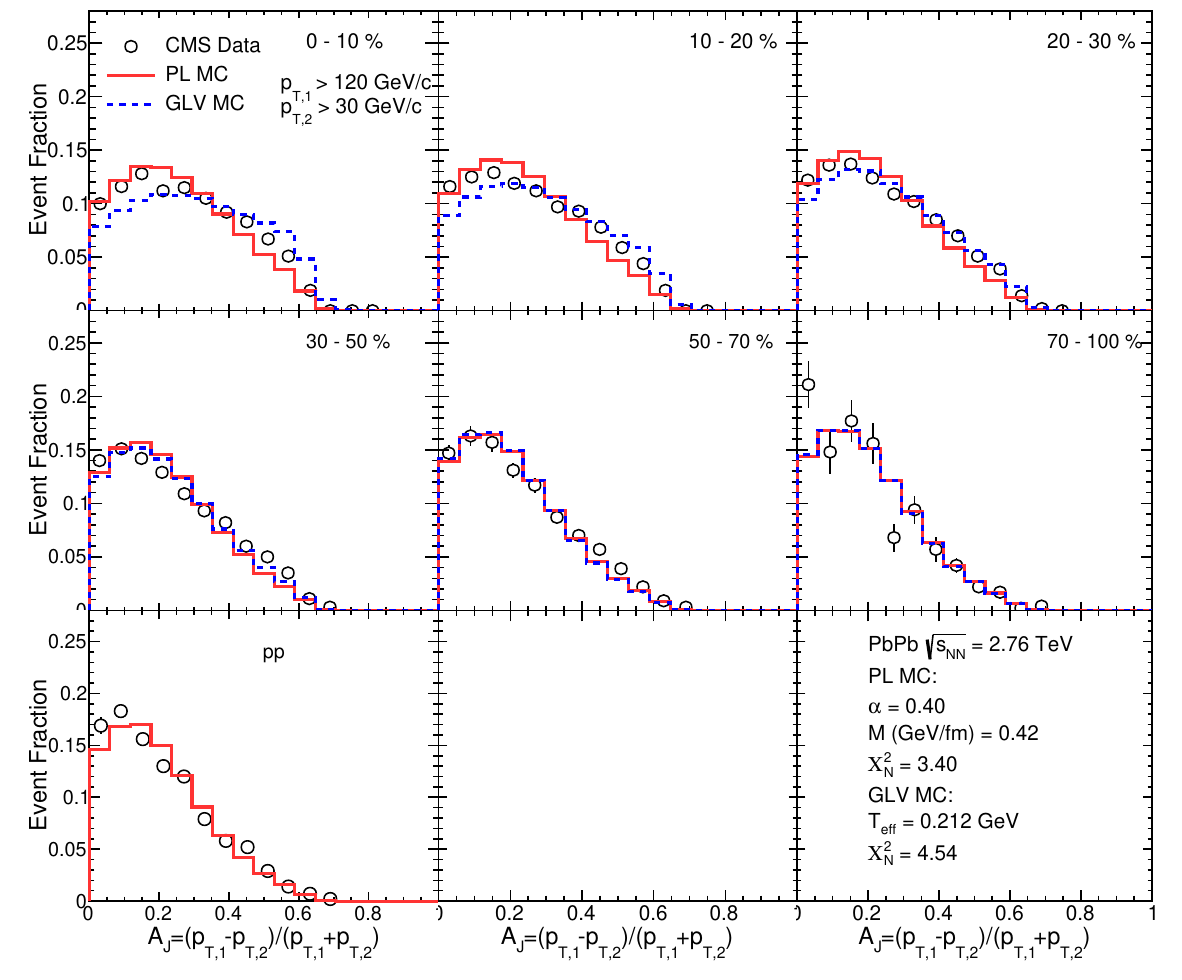}
  \caption{(Color online) Dijet asymmetry distribution in different centrality windows 
    in PbPb collisions at $\sqrt{s_{NN}}$ = 2.76 TeV measured by
    the CMS experiment~\cite{CMS:2012ulu} compared with our calculations
    using power law and GLV energy loss. 
    The parameters are $\alpha=0.40$ and $M=0.42$ GeV/fm for PL
    and $T_{\rm eff} = 212$ MeV for GLV.}
  \label{Fig:DiJetAsymCent}
\end{figure*}

Figure~\ref{Fig:LatticeEOS} (right) shows transverse size of the system as a function of
$N_{\rm Part}$ of the collisions. The dashed blue line shows the transverse size, $R(\tau_0)$, at the
initial time $\tau_{0}$ while long dashed red line shows the effective transverse size,
$R^{\rm eff} = R(\tau_0) + a_{t}\tau_{F}^{3}/6$, 
obtained by integrating the $R(\tau$) over the time evolution of the system.
Table~\ref{Tab:ParDetails} shows the average number of participants,
the mean transverse radius of the medium and the mean values of effective
path-lengths for the two back-to-back partons in several classes of collision
centralities for PbPb collisions at $\sqrt{s_{NN}}$ = 2.76 TeV and 5.02 TeV
along with the values of $R^{\rm eff}$.
In the peripheral collisions, the transverse plane looks like an ellipse.
Since the transverse expansion is faster in the short axis direction,
the transverse plane would finally lead to a near circular shape.

\begin{table}[h] %add [H] placement to break table across pages
  \centering
  \caption{\label{Tab:ParDetails} The average number of participants ({\NPartAv}),
    the mean transverse radius of the medium ({\RAv}) and
    the mean values of effective path-lengths for the two back-to-back partons
    ({\LAAv}, {\LBAv}) in several classes of collision centralities for PbPb
    collisions at $\sqrt{s_{NN}}$ = 2.76 TeV (I) and at 5.02 TeV (II).}
    \begin{tabular}{ccccccccc}
      \hline
      Cent($\%$)\,\,       &\multicolumn{2}{c}{{\NPartAv}\,\,\,}      &\multicolumn{2}{c}{{\RAv}(fm)\,\,\,} &\multicolumn{2}{c}{{\LAAv}(fm)\,\,\,}   &\multicolumn{2}{c}{{\LBAv}(fm)\,\,\,}  \\ \hline
                       &{\SNNA} &{\SNNB}                    &{\SNNA} &{\SNNB}          &{\SNNA} &{\SNNB}               &{\SNNA} &{\SNNB} \\ \hline
      0-10             &355     &359                        &8.82  &9.20               &7.55   &7.88                   &7.70   &8.03                  \\
      10-20            &261     &264                        &7.55  &7.86               &6.45   &6.74                   &6.57   &6.86                  \\
      20-30            &187     &189                        &6.40  &6.69               &5.48   &5.70                   &5.59   &5.81                  \\
      30-40            &130     &131                        &5.36  &5.58               &4.59   &4.78                   &4.67   &4.88                  \\
      40-50            &86      &87                         &4.42  &4.59               &3.78   &3.93                   &3.86   &4.01                  \\
      50-60            &54      &54                         &3.52  &3.67               &3.01   &3.12                   &3.07   &3.18                  \\
      60-70            &31      &31                         &2.69  &2.79               &2.30   &2.39                   &2.34   &1.43                  \\
      70-100           &9       &9                          &1.43  &1.45               &1.23   &1.40                   &1.25   &1.43                  \\ \hline
                                                                                                                                    
      0-100            &113     &114                        &5.01  &5.23               &4.29   &4.45                   &4.37   &4.54                  \\ \hline
    \end{tabular}
\end{table}

%The transverse size of the medium is decided by centrality of collisions as 
%\begin{equation}
%  R = R_{A}\left(\frac{N_{\rm part}}{2A}\right)^{\beta} .
%\label{eqcentPart}
%\end{equation}
%Here, $R_{A}$ and $A$ are the radius and mass number of the colliding nuclei
%and $N_{\rm part}$ is the number of participant in a centrality class~\cite{Kumar_2019}.
%The value of $\beta=0.55$ is used in our calculations.

Figure~\ref{Fig:DiJetDiagram} shows a snapshot of the transverse cross section of expanding QGP
medium of size $R(\tau)$ at an instant $\tau$ between initial time $\tau_{0}$ and kinetical freeze out
time. As seen in Fig.~\ref{Fig:DiJetDiagram} two back-to-back jets produced at radial distance $r$
could traverse different path-lengths ($L_{1}$ and $L_{2}$) in the medium. The value of freeze
out temperature is used as 0.140 GeV in our calculations.

The position $r$ is generated using a 2$\pi r$ distribution.
The directions of the jets given by $\phi$ in dijet or $\gamma$-jet pairs are
generated randomly between 0 to 2$\pi$ within the transverse cross section
of the medium.
The path-lengths $L_1$ and $L_2$ can be expressed as
\begin{eqnarray}
  \centering
  L_{1} &= &\sqrt{(R^{\rm eff})^{2}-r^{2}~\sin^2(\phi)} - r~\cos(\phi)   \nonumber \\  
  L_{2} &= &\sqrt{(R^{\rm eff})^{2}-r^{2}~\sin^2(\pi+\phi)} - r~\cos(\pi+\phi).
\end{eqnarray}
The path-lengths are kept smaller than the life time of the medium.
The energy loss $\Delta E$ of jets 1 and 2 can be calculated as
\begin{equation}
  \Delta E_1 = \frac{dE}{dx} \times L_1, \hspace{1cm}  \Delta E_2 = \frac{dE}{dx} \times L_2. 
\end{equation}
Note that $dE/dx$ is also path-lengths ($L_1$, $L_2$) dependent in case of GLV. 
If \ptx is the generated transverse momentum of the jet then
the final transverse momentum of the jet in dijet
or $\gamma$-jet pair will be
\begin{equation}
\centering
   \pta =  \ptx - \Delta E_1, \nonumber {\hspace{1cm}}
   \ptb =  \ptx - \Delta E_2.    
\end{equation}
The transverse momentum imbalance can be quantified using various asymmetry parameters
defined as 
\begin{eqnarray}
  \centering
  A_{J} & = \frac{\pta-\ptb}{\pta+\ptb}, \nonumber \\
  X_{J} & = \frac{\ptb}{\pta} .  
\end{eqnarray}
Both these measures are used by experimental groups and these are related to each other as
$A_J=(1-X_J)/(1+X_J)$.
    
The asymmetry parameter of the $\gamma$-jet pairs can be calculated
as follows
\begin{equation}
  X_{J\gamma} = \frac{p_T^{\rm Jet}}{p_T^{\rm \gamma}}.
\end{equation}
For the $\gamma$-jet pairs, no energy loss is assigned for the photon as it
does not participate in the strong interaction. 

%Other than the energy loss, the jet \ptx will also be modified due to the
%experimental resolution of jet reconstruction.
%For CMS data we use the \ptx dependent resolution factor used by the
%CMS collaboration~\cite{CMS:2017eqd}. The typical values for \ptx resolution are
%$\approx$20$\%$ for low \ptx jets.
%First, the asymmetry distribution in pp collisions are reproduced.
%The additional broadening in PbPb collisions can then be attributed to
%energy loss in the medium but it may not be free from experimental effect.
%The measurements from ATLAS are claimed to account for underlying 
%background and experimental effects~\cite{ATLAS:2017xfa,ATLAS:2022zbu}.

\begin{figure*}
  \includegraphics[width=0.99\textwidth]{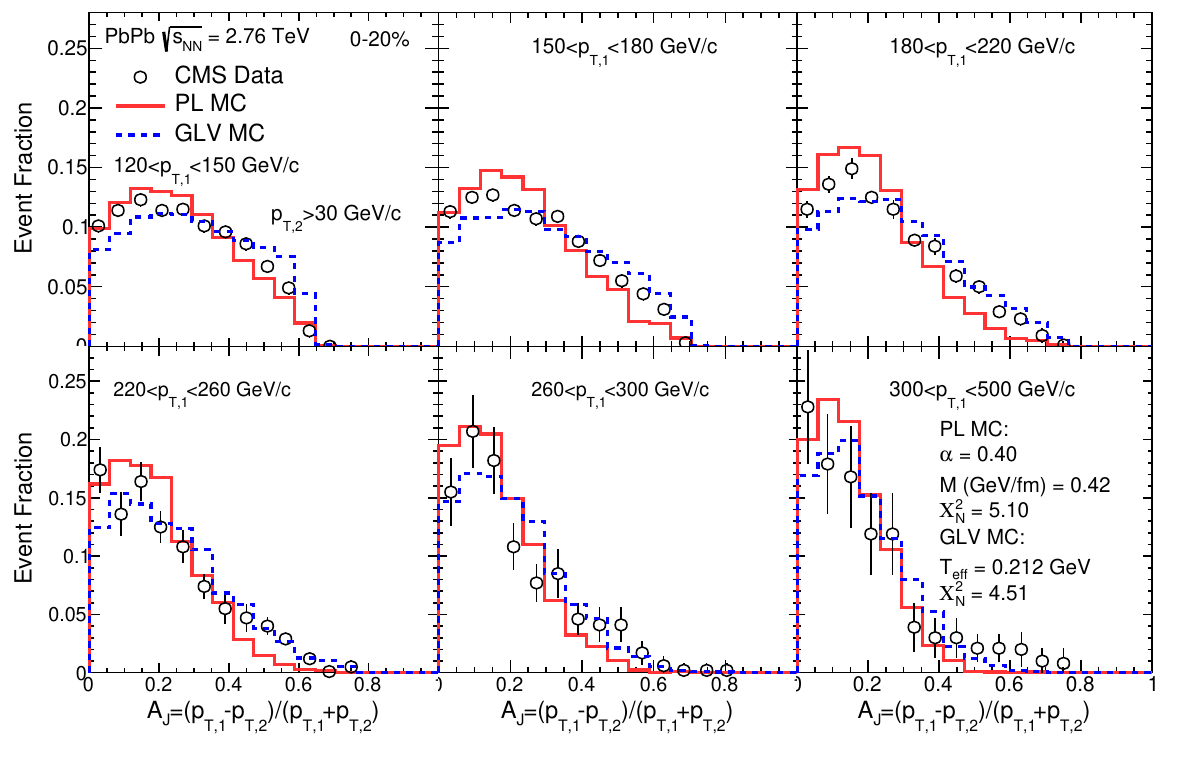}
  \caption{(Color online) Dijet asymmetry in different \ptx windows of 
    of leading jet in PbPb collisions at $\sqrt{s_{NN}}$ = 2.76 TeV measured by
    the CMS experiment~\cite{CMS:2012ytf} in 0-20$\%$ centrality
    bin compared with our calculations
    using power law and GLV energy loss. 
    The parameters are $\alpha=0.40$ and $M=0.42$ GeV/fm for PL
    and $T_{\rm eff} = 212$ MeV for GLV.}
  \label{Fig:DiJetAsymPt}
\end{figure*}

\begin{figure*}
  \includegraphics[width=0.99\textwidth]{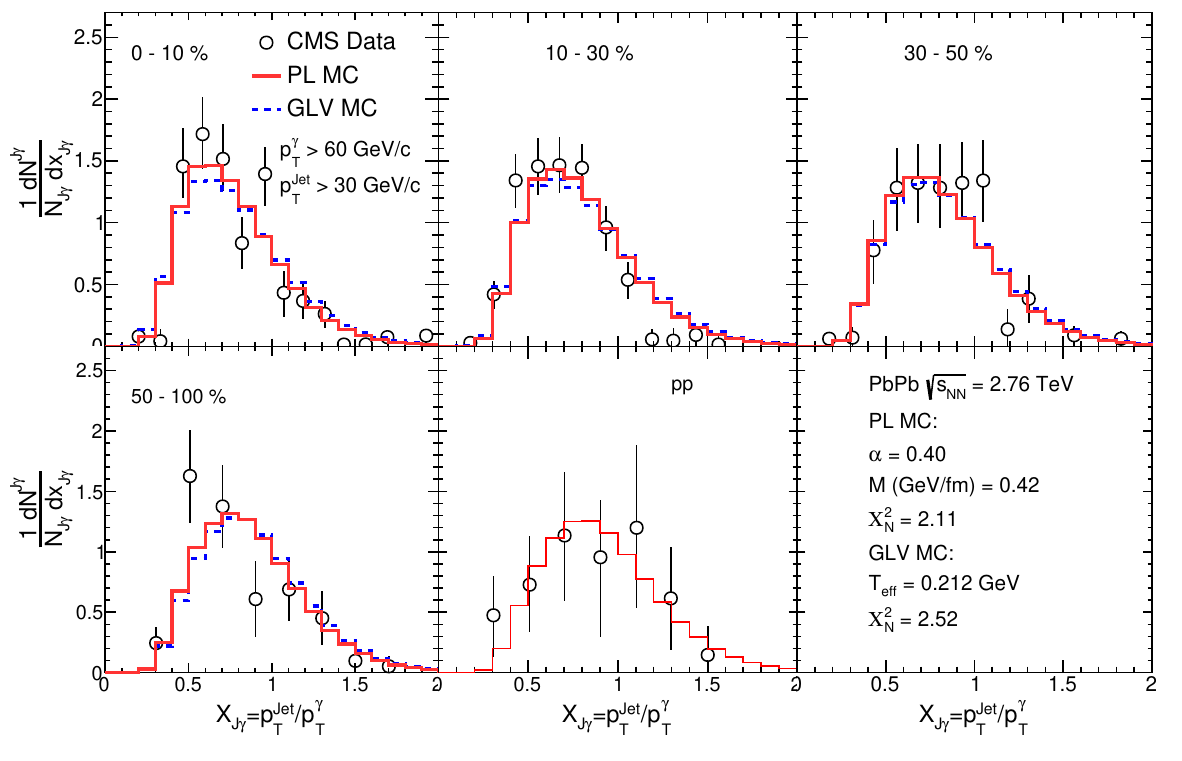}
  \caption{(Color online) The $\gamma$-jet asymmetry for different centrality regions
    in PbPb collisions at $\sqrt{s_{NN}}$ = 2.76 TeV as measured by CMS
    experiment~\cite{CMS:2012ytf} along with the calculations
    using power law and GLV energy loss. 
    The parameters are $\alpha=0.40$ and $M=0.42$ GeV/fm for PL
    and $T_{\rm eff} = 212$ MeV for GLV.}
  \label{Fig:JetAsymGammaJetCent}
\end{figure*}

\begin{figure*}
  \includegraphics[width=0.99\textwidth]{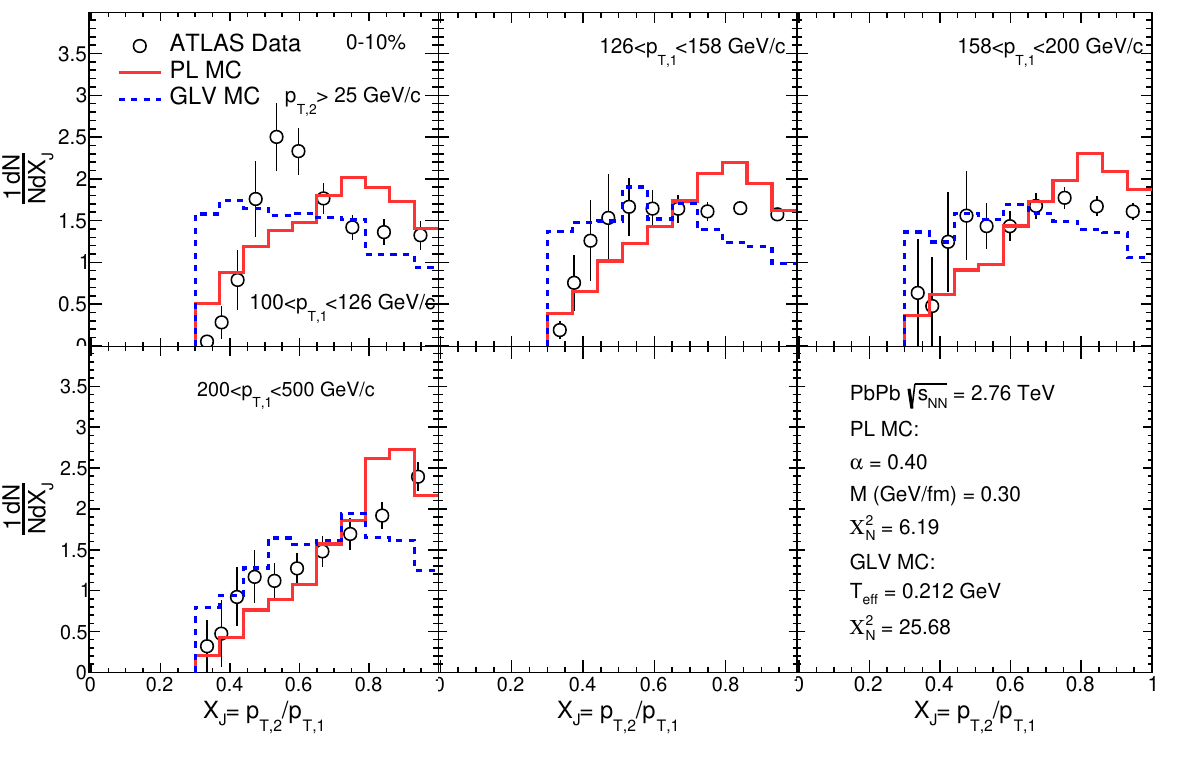}
  \caption{(Color online) The unfolded Di-jet asymmetry ($X_{\rm J}$) for different $p_{\mathrm{T}}$ bins
    in PbPb collisions at $\sqrt{s_{NN}}$ = 2.76 TeV and in pp collisions at $\sqrt{s}$ = 2.76 TeV
    measured by ATLAS experiment~\cite{ATLAS:2017xfa} along with the calculations
    using power law and GLV energy loss. 
    The parameters used are $\alpha = 0.40$ and $M = 0.30$ GeV/fm for PL
    and $T_{\rm eff} = 212$ MeV for GLV.
  }
  \label{Fig:ATLASXJPt276TeV}
\end{figure*}

\begin{figure*}
  \includegraphics[width=0.99\textwidth]{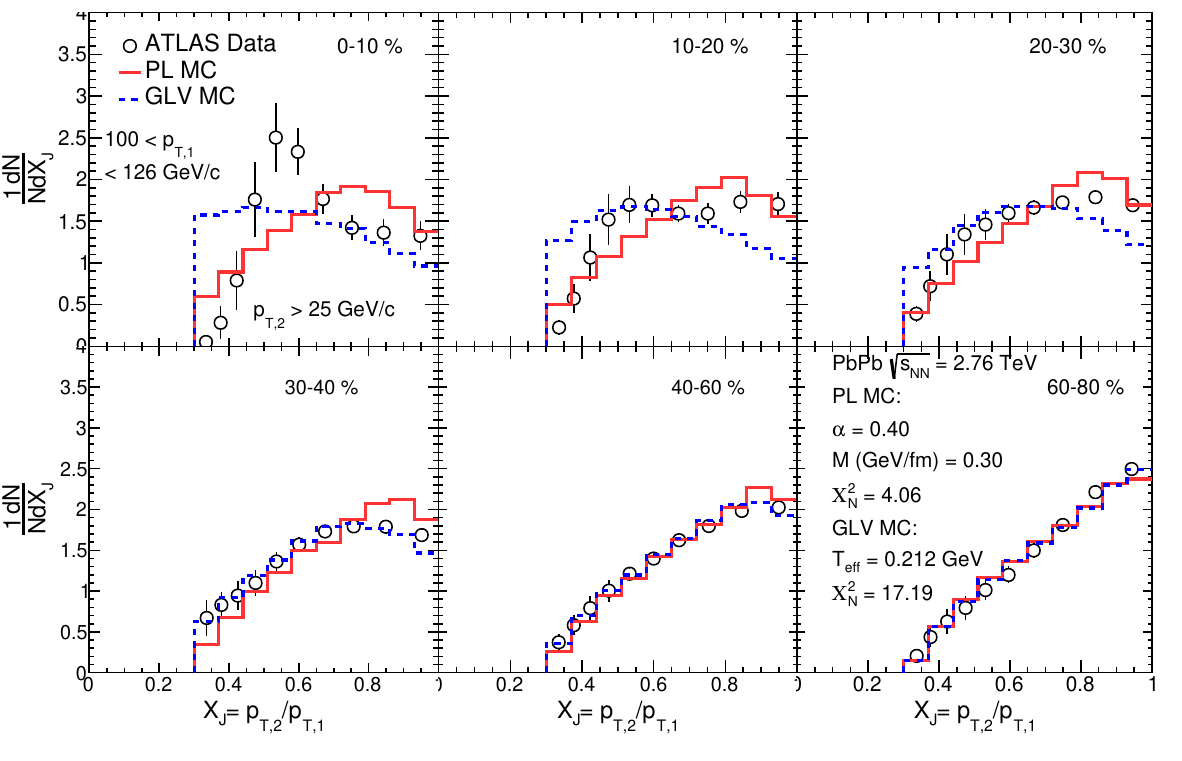}
  \caption{(Color online) The unfolded Di-jet asymmetry ($X_{\rm J}$) for different centrality windows
    in PbPb collisions at $\sqrt{s_{NN}}$ = 2.76 TeV and in pp collisions at $\sqrt{s}$ = 2.76 TeV
    measured by ATLAS experiment~\cite{ATLAS:2017xfa} along with the calculations
    using power law and GLV energy loss. 
    The parameters used are $\alpha = 0.40$ and $M = 0.30$ GeV/fm for PL
    and $T_{\rm eff} = 212$ MeV for GLV.
  }
  \label{Fig:ATLASXJCent276TeV}
\end{figure*}

\begin{figure*}
  \includegraphics[width=0.99\textwidth]{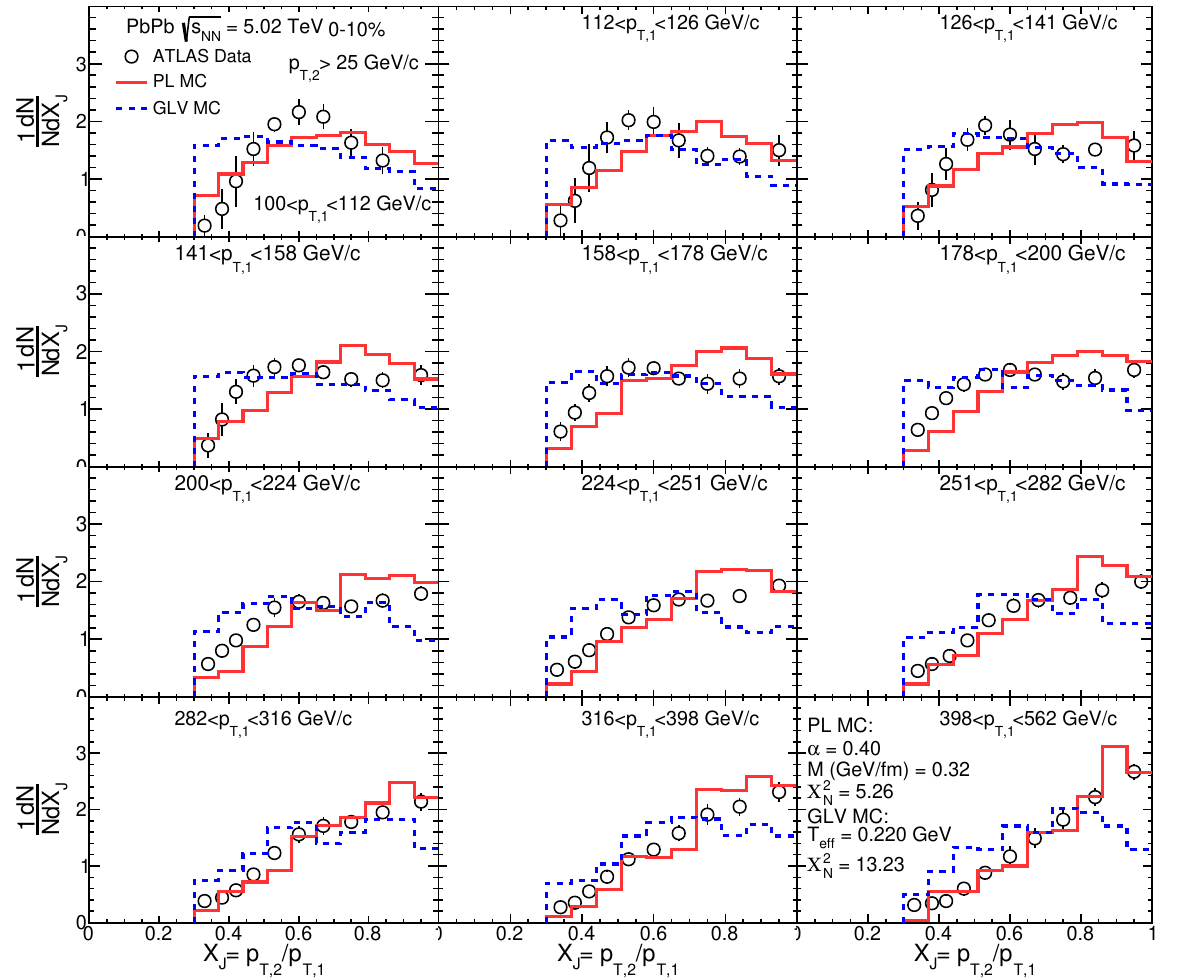}
  \caption{(Color online) The unfolded Di-jet asymmetry ($X_{\rm J}$) in different $p_{\mathrm{T}}$ bins
    in PbPb collisions at $\sqrt{s_{NN}}$ = 5.02 TeV in the most central 0-10$\%$ collisions
    measured by ATLAS experiment~\cite{ATLAS:2022zbu} along with the calculations
    using power law and GLV energy loss.
    The parameters used are $\alpha = 0.40$ and $M = 0.32$ GeV/fm for PL
    and $T_{\rm eff} = 220$ MeV for GLV.
  }
  \label{Fig:ATLASXJPtBinsCent010at502TeV}
\end{figure*}

\begin{figure*}
  \includegraphics[width=0.99\textwidth]{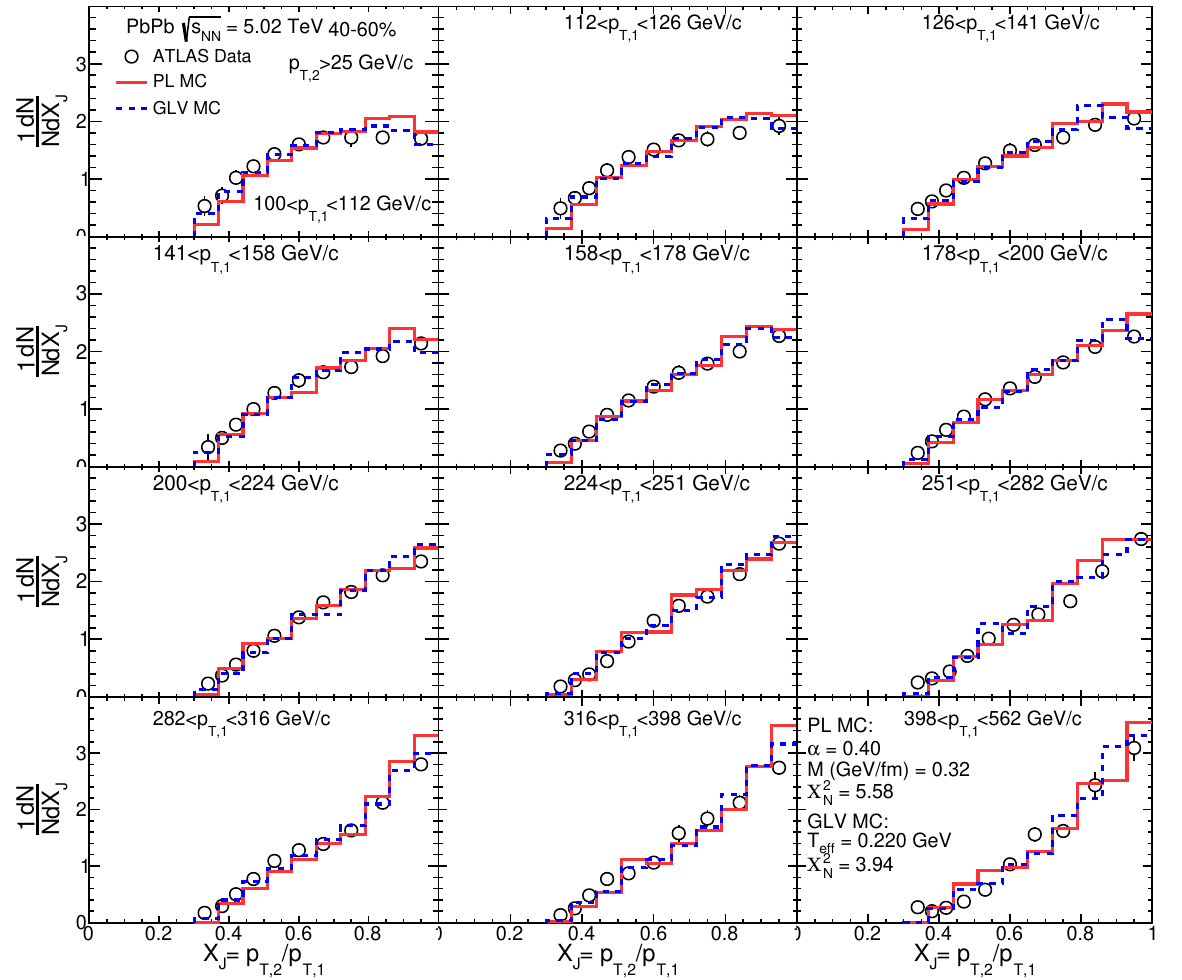}
  \caption{(Color online) The unfolded Di-jet asymmetry (X$_{\rm J}$) in different $p_{\mathrm{T}}$ bins
    in PbPb collisions at $\sqrt{s_{NN}}$ = 5.02 TeV in the peripheral 40-60$\%$ collisions
    measured by ATLAS experiment~\cite{ATLAS:2022zbu} along with the calculations
    using power law and GLV energy loss.
    The parameters used are $\alpha = 0.40$ and $M = 0.32$ GeV/fm for PL
   and $T_{\rm eff} = 220$ MeV for GLV.
  }
  \label{Fig:ATLASXJPtBinsCent4060at502TeV}
\end{figure*}

 We generate the \ptx of jets using the measured \ptx distribution and the imbalance
in \ptx is also obtained using the measured pp data.
Thus, the generated \ptx takes into account both the vacuum effect and any experimental effect in pp. 
For CMS data (Fig.~\ref{Fig:DiJetAsymCent}, Fig.~\ref{Fig:DiJetAsymPt} and
Fig.~\ref{Fig:JetAsymGammaJetCent}) we use the \ptx
dependent resolution factor used by the CMS collaboration~\cite{CMS:2017eqd}.
The \ptx broadening is taken into account using a Gaussian distribution with width given by
experimental resolution multiplied by a constant scale factor ($\approx$1.7) such that  asymmetry
distribution in pp collisions are reproduced.
The additional broadening in PbPb collisions can then be attributed to energy loss in the medium
but it may not be free from experimental effect depending on the experimental setup.
The measurements from ATLAS are claimed to account for underlying background and experimental
effects~\cite{ATLAS:2017xfa,ATLAS:2022zbu}. For the ATLAS data (Fig.~\ref{Fig:ATLASXJPt276TeV},
Fig.~\ref{Fig:ATLASXJCent276TeV}, Fig.~\ref{Fig:ATLASXJPtBinsCent010at502TeV} and
Fig.~\ref{Fig:ATLASXJPtBinsCent4060at502TeV}) the \ptx of one jet is generated using measured
\ptx distribution and the \ptx of the other jet is obtained using pp asymmetry distributions.

We also calculate the nuclear modification factor $R_{\rm AA}$ for jets.
ATLAS and CMS experiments measured the $R_{\rm AA}$ for the jets at $\sqrt{s_{NN}} = $ 2.76 TeV
and 5.02 TeV respectively. The dijet events are generated with \ptx distributed
as per Eq.~\ref{EQhage} and the \ptx of the jets are filled in 
a histogram $N_{\rm pp}(p_T)$. The modified transverse momenta of the two jets
after loosing energy in the medium are filled in histogram $N_{\rm PbPb}(p_T)$.
The ratio of the two histograms gives the nuclear modification factor

\begin{equation}
R_{\rm AA} = \frac{N_{\rm PbPb}(p_T)}{N_{\rm pp}(p_T)}.
\end{equation}
For all the analysis, the experimental cuts are implemented which are
$\pta \geq 120$ GeV/$c$ and $\ptb \geq 30$ GeV/$c$ for CMS experiment and
$\pta \geq 100$ GeV/$c$ and $\ptb \geq 25$ GeV/$c$ for ATLAS experiment.
$p_T^{\rm Jet} \geq 30$ GeV/$c$ and $p_T^{\gamma} \geq 60$ GeV/$c$
for CMS $\gamma$-Jet measurement. In addition, $\Delta\phi\geq$2$\pi$/3 for
CMS experiment and $\Delta\phi\geq$7$\pi$/8 for ATLAS experiment,
where $\phi$ is the azimuthal angle. The calculations for R$_{\rm AA}$ does
not have the $\Delta\phi$ cut in accordance to the experimental measurements.

\begin{figure*}
  \includegraphics[width=0.99\textwidth]{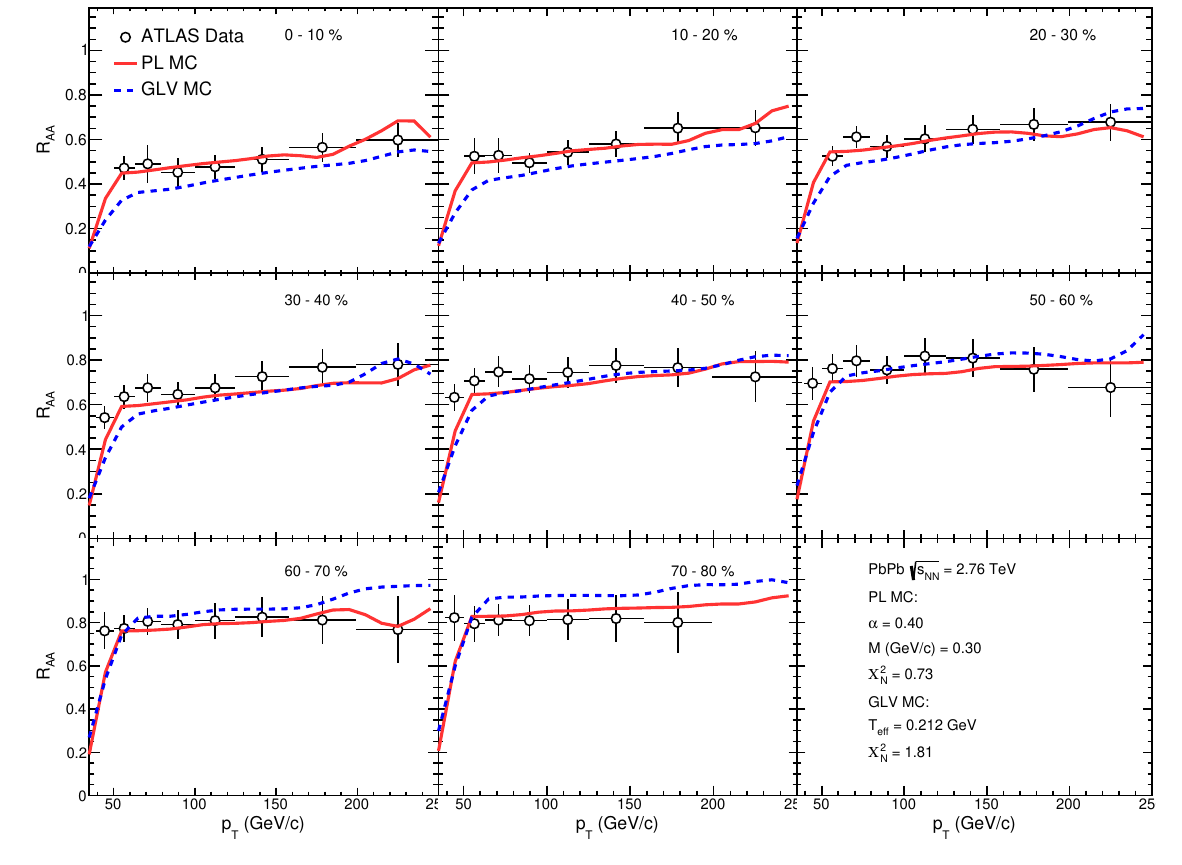}
  \caption{(Color online)Jet $R_{AA}$ as  a function of jet \ptx in several collision centrality bins
    in PbPb collisions at $\sqrt{s_{NN}}$ = 2.76 TeV measured by the ATLAS experiment~\cite{ATLAS:2014ipv}.
    The measurements are compared calculations
    using power law and GLV energy loss.
    The parameters used are $\alpha = 0.40$ and $M = 0.30$ GeV/fm for PL
    and $T_{\rm eff} = 212$ MeV for GLV.
  }
  \label{Fig:JetRAAPt_ATLAS_276TeV}
\end{figure*}

\begin{figure*}
  \includegraphics[width=0.99\textwidth]{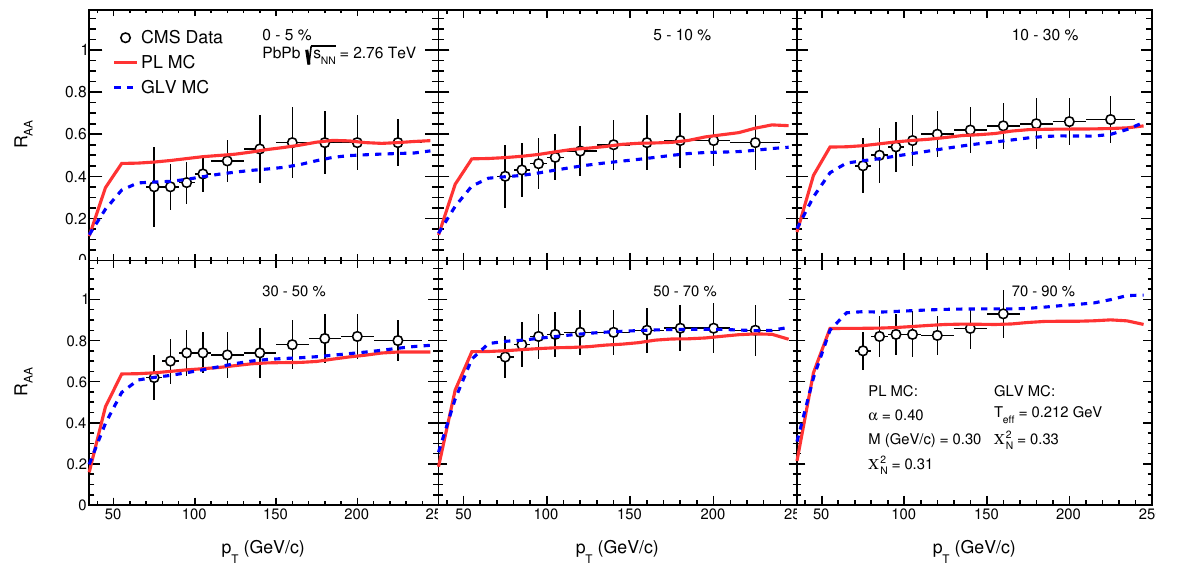}
  \caption{(Color online)Jet $R_{AA}$ as  a function of jet \ptx in several collision centrality bins
    in PbPb collisions at $\sqrt{s_{NN}}$ = 2.76 TeV measured by the CMS experiment~\cite{CMS:2016uxf}.
    The measurements are compared calculations
    using power law and GLV energy loss.
    The parameters used are $\alpha = 0.40$ and $M = 0.30$ GeV/fm for PL
    and $T_{\rm eff} = 212$ MeV for GLV.
  }
  \label{Fig:IncJetRAAPt_CMS_276TeV}
\end{figure*}

\begin{figure*}
  \includegraphics[width=0.99\textwidth]{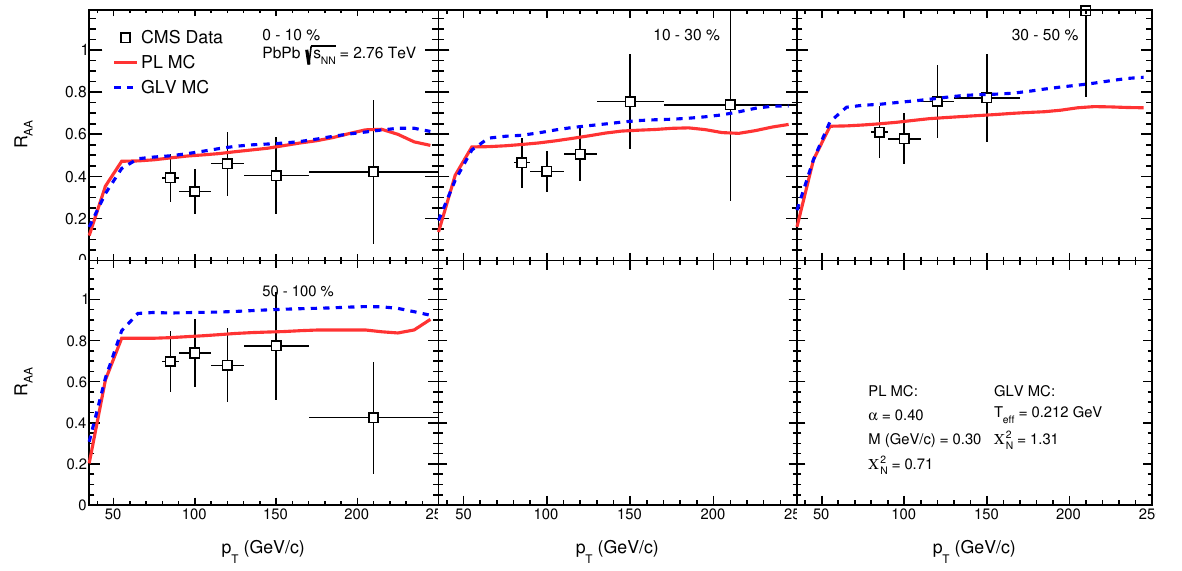}
  \caption{(Color online)b-Jet $R_{AA}$ as  a function of jet \ptx in several collision centrality bins
    in PbPb collisions at $\sqrt{s_{NN}}$ = 2.76 TeV measured by the CMS experiment~\cite{CMS:2013qak}.
    The measurements are compared calculations
    using power law and GLV energy loss.
    The parameters used are $\alpha = 0.40$ and $M = 0.30$ GeV/fm for PL
    and $T_{\rm eff} = 212$ MeV for GLV.
  }
  \label{Fig:bJetRAAPt_CMS_276TeV}
\end{figure*}

%\begin{figure*}
%  \includegraphics[width=0.99\textwidth]{Fig_CMSXJPlots_502TeV_Cent_Band.pdf}
%  \caption{(Color online) Dijet asymmetry distribution in different centrality windows 
%    in PbPb collisions at $\sqrt{s_{NN}}$ = 5.02 TeV measured by
%    the CMS experiment for inclusive and b jets~\cite{CMS:2018dqf} compared with our calculations
%    using power law and GLV energy loss. The parameters are $\alpha=0.40$ and $M=0.45$ GeV/fm for PL
%    and $T_{\rm eff} = 220$ MeV for GLV.}
%  \label{Fig:DiJetAsymCent}
%\end{figure*}

\begin{figure*}
  \includegraphics[width=0.85\textwidth]{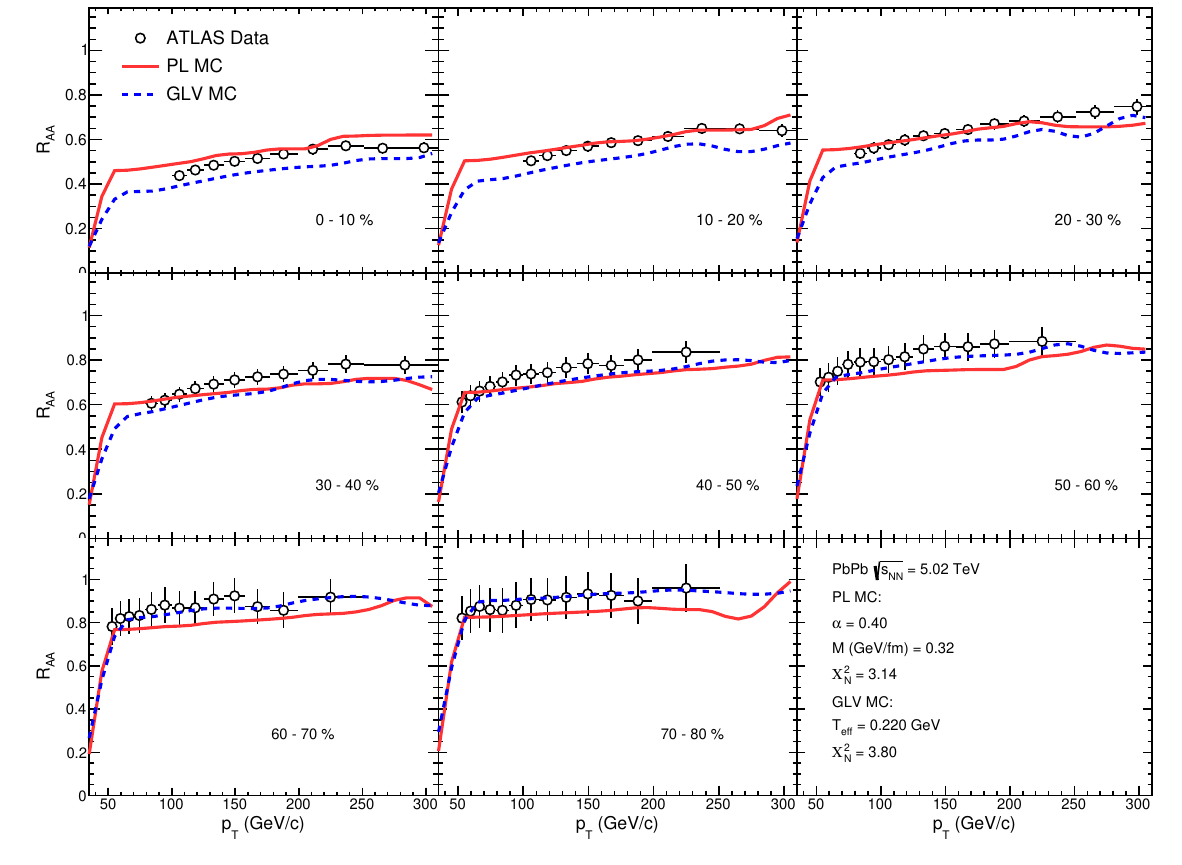}
  \caption{(Color online) Jet $R_{AA}$ as  a function of jet \ptx in several collision centrality bins
    in PbPb collisions at $\sqrt{s_{NN}}$ = 5.02 TeV measured by the ATLAS experiment~\cite{ATLAS:2018gwx}.
    The measurements are compared with calculations
    using power law and GLV energy loss.
    The parameters used are $\alpha = 0.40$ and $M = 0.32$ GeV/fm for PL
    and $T_{\rm eff} = 220$ MeV for GLV.
  }
  \label{Fig:JetRAAPt_ATLAS_502TeV}
\end{figure*}

\begin{figure*}
  \includegraphics[width=0.99\textwidth]{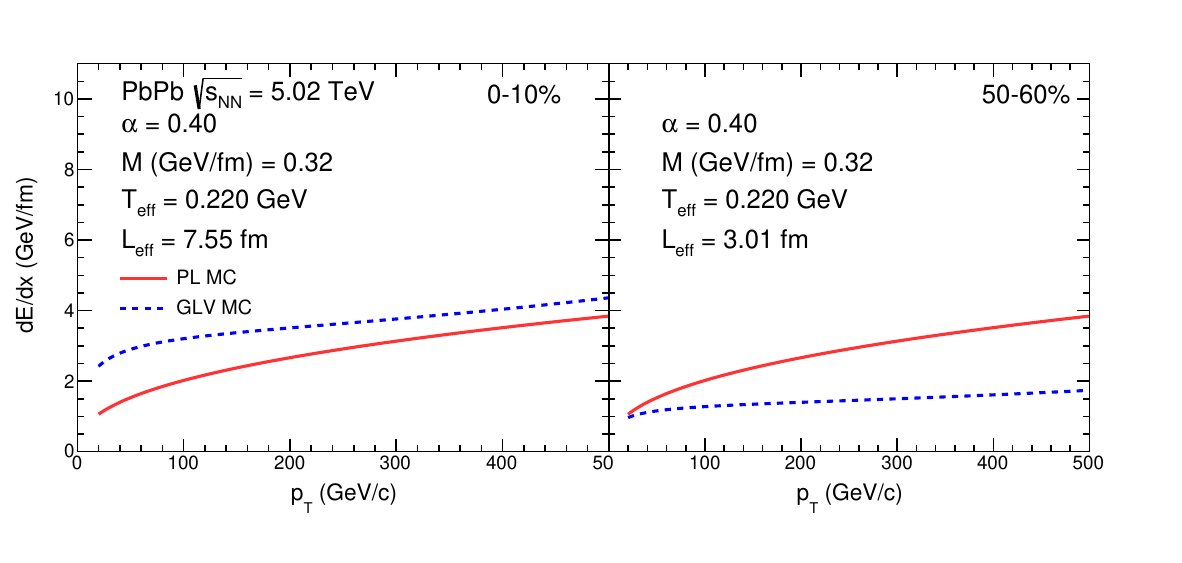}
  \caption{(Color online) The energy loss per unit lenght, dE/dx, as a function of jet \ptx in
    central (0-10$\%$) and peripheral (50-60$\%$) collisions for both PL and GLV
    models.
      }
  \label{Fig:dEdxComp_MM32}
\end{figure*}

%\begin{figure*}
%  \includegraphics[width=0.99\textwidth]{Fig_XJ_Z0Jet_Centrality.pdf}
%  \caption{(Color online) (Color online) Jet asymmatry as a function of collision centrality
%    in Z$^{0}$ + Jet events as measured by CMS experiment. The data is compared with our
%    calculations.}
%  \label{Fig:JetAsymZ0JetCent}
%\end{figure*}

%\begin{table}[h] %add [H] placement to break table across pages
%\centering
%\caption{\label{Tab:ParValues} The values of parameters $\alpha$ and $M$ extracted from CMS and ATLAS measurements at
%  {\SNNO} and {\SNNT} for Power Law MC method. The value of $T_{\rm eff}$ used for GLV MC is also shown.}
%\begin{tabular}{lccc|cc}
%\hline
%      &\multicolumn{3}{c}{{\SNNO}}                         &\multicolumn{2}{|c}{{\SNNT}}  \\
%\hline
%      %&CMS    &\multicolumn{2}{c}{ATLAS}    &\multicolumn{1}{c}{ATLAS} &ATLAS                  \\ \hline
%      &CMS     &\multicolumn{2}{c}{ATLAS}    &\multicolumn{2}{|c}{ATLAS}                   \\
%\hline
%            & ($A_{J}$, $X_{J}^{\gamma}$)  & $X_{J}$  &$R_{\rm AA}$       &$X_{J}$      &$R_{\rm AA}$     \\
%\hline
% $\alpha$              &0.40            &0.40    &0.40              &0.40        &0.40   \\      
% $M$ (GeV/fm)          &0.42            &0.30    &0.30              &0.32        &0.32  \\      
% $T_{\rm eff}$ (GeV)     &0.212           &0.212   &0.212             &0.220       &0.220  \\
%\hline
%\end{tabular}
%\end{table}

\begin{table}[h] %add [H] placement to break table across pages
\centering
\caption{\label{Tab:ParValues} The values of parameters $\alpha$ and $M$ extracted from CMS and ATLAS measurements at
  {\SNNO} and {\SNNT} for Power Law MC method. The value of $T_{\rm eff}$ used for GLV MC is also shown.}
\begin{tabular}{lcc|c}
\hline
      &\multicolumn{2}{c}{{\SNNO}}                         &\multicolumn{1}{|c}{{\SNNT}}  \\
\hline
      %&CMS    &\multicolumn{2}{c}{ATLAS}    &\multicolumn{1}{c}{ATLAS} &ATLAS                  \\ \hline
      &CMS     &\multicolumn{1}{c}{ATLAS}    &\multicolumn{1}{|c}{ATLAS}                   \\
\hline
            & ($A_{J}$, $X_{J}^{\gamma}$)  &($X_{J}$, $R_{\rm AA}$)       &($X_{J}$,$R_{\rm AA}$)     \\
\hline
 $\alpha$              &0.40             &0.40                   &0.40   \\      
 $M$ (GeV/fm)          &0.42             &0.30                   &0.32  \\      
 $T_{\rm eff}$ (GeV)     &0.212            &0.212                  &0.220  \\
\hline
\end{tabular}
\end{table}

\section{Results and discussions}
\label{Sec:ResultsAndDiss}
We have used two models, Power Law (PL) and GLV to describe the experimental data.
While energy loss in case of PL depends on $L$, it depends on
$L^2$ for GLV case. In former case the energy loss depends on jet energy
by power law and in later case it is logarithmic in jet energy.
CMS data of dijet and $\gamma$-jet asymmetry in PbPb collisions at $\sqrt{s_{NN}}$ = 2.76 TeV
has been used in the analysis.
Also ATLAS data of dijet asymmetry and $R_{\rm AA}$ in PbPb collisions at two energies
$\sqrt{s_{NN}}$ = 2.76 TeV and 5.02 TeV have been used. 
We had aimed to keep two sets of parameters for two collision energies but we did
vary them for different experiments and discuss the underlying
reason behind it. 

Figure~\ref{Fig:DiJetAsymCent} shows
dijet asymmetry distribution in different centrality windows 
in PbPb collisions at $\sqrt{s_{NN}}$ = 2.76 TeV measured by
the CMS experiment~\cite{CMS:2012ulu} compared with our calculations
using power law and GLV energy loss. 
The parameters obtained for CMS are $\alpha=0.40$ and $M=0.42$ GeV/fm for PL.
The GLV formalism also gives reasonable description of the data using the extracted value of
the effective temperature $T_{\rm eff} = $ 212 MeV.
The distributions of $A_J$ become narrower as we move from central to peripheral
collisions, the feature is well produced by both PL and GLV by
given parameters with both GLV having slightly higher $\chi^2/ndf$ values.
  Figure~\ref{Fig:DiJetAsymPt} shows dijet asymmetry in different \ptx windows 
of leading jet in PbPb collisions at $\sqrt{s_{NN}}$ = 2.76 TeV measured by
the CMS experiment~\cite{CMS:2012ytf} in 0-20$\%$ centrality
bin compared with our calculations using power law and GLV energy loss. 
%The parameters are $\alpha=0.46$ and $M=0.57$ GeV/fm for PL
%and $T_{\rm eff} = 300$ MeV for GLV.
The distribution becomes narrower as the \ptx of the jets increases, a trend
correctly produced by both PL and GLV with above parameters with both models
 having similar $\chi^2/ndf$ values.
 Figure~\ref{Fig:JetAsymGammaJetCent} shows
the $\gamma$-jet asymmetry for different centrality regions
in PbPb collisions at $\sqrt{s_{NN}}$ = 2.76 TeV as measured by CMS
experiment~\cite{CMS:2012ytf} along with the calculations
using power law and GLV energy loss. 
%The parameters are $\alpha=0.46$ and $M=0.57$ GeV/fm for PL
%and $T_{\rm eff} = 300$ MeV for GLV.
The same parameters used to describe dijet calculations
are able to explain the the $\gamma$-jet asymmetry for both PL and GLV within
experimental errors. 
Since the broadening of the asymmetry distribution in case of all CMS data
may also include the effect of underlying events,
the values of the parameters can be considered as upper limit.

The measurements from ATLAS account for underlying
background and experimental effects~\cite{ATLAS:2017xfa,ATLAS:2022zbu} and thus the
shapes of the asymmetry distributions is supposed to arise purely from
the energy loss effects.
Figure~\ref{Fig:ATLASXJPt276TeV} shows
the unfolded Di-jet asymmetry ($X_{\rm J}$) for different $p_{\mathrm{T}}$ bins
%in PbPb collisions at $\sqrt{s_{NN}}$ = 2.76 TeV and in pp collisions at $\sqrt{s}$ = 2.76 TeV
in PbPb collisions at $\sqrt{s_{NN}}$ = 2.76 TeV measured by ATLAS
experiment~\cite{ATLAS:2017xfa} along with the calculations
using power law and GLV energy loss. 
The parameters obtained for ATLAS data at  $\sqrt{s_{NN}}$ = 2.76 TeV
are $\alpha = 0.40$ and $M = 0.30$ GeV/fm for PL and $T_{\rm eff} = 212$ MeV for GLV. 
Except the distribution in the lowest $p_{\mathrm{T}}$ window, the data is well
described by PL. The lower values of $\chi^2/ndf$ for the case of PL show that 
it describes the data much better as compared to GLV.
It can be noted that, the ATLAS data requires smaller values of $M$ 
as ATLAS reportedly accounted for experimental effects affecting the broadening of
$X_{\rm J}$ distributions. 
Figure~\ref{Fig:ATLASXJCent276TeV} shows the 
unfolded Di-jet asymmetry ($X_{\rm J}$) for different centrality windows
%in PbPb collisions at $\sqrt{s_{NN}}$ = 2.76 TeV and in pp collisions at $\sqrt{s}$ = 2.76 TeV
in PbPb collisions at $\sqrt{s_{NN}}$ = 2.76 TeV 
measured by ATLAS experiment~\cite{ATLAS:2017xfa} along with the calculations
using power law and GLV energy loss. 
%The parameters obtained are $\alpha = 0.42$ and $M = 0.50$ GeV/fm for PL
%and $T_{\rm eff} = 300$ MeV for GLV.
Here also, the lower values of $\chi^2/ndf$ for the case of PL show that 
it describes the data much better as compared to GLV.
{\color{black}
  Figure~\ref{Fig:JetRAAPt_ATLAS_276TeV} and Fig.~\ref{Fig:IncJetRAAPt_CMS_276TeV} shows the
  jet $R_{AA}$ as a function of jet \ptx in several collision centrality bins in PbPb collisions
  at $\sqrt{s_{NN}}$ = 2.76 TeV measured by the ATLAS and CMS
  experiments~\cite{ATLAS:2014ipv,CMS:2016uxf} respectively. The measurements are compared with the
  calculations using power law and GLV energy loss.
}
%The parameters obtained are $\alpha = 0.38$ and $M = 0.46$ GeV/fm for PL
%and $T_{\rm eff} = 300$ MeV for GLV.
The $R_{AA}$ value slowly inches towards one as one moves to higher $p_T$ but, remains
below one. The $R_{AA}$ approaches one as one moves from central to peripheral
collisions. The overall dependence of $R_{AA}$ on $p_T$ is correctly
reproduced by both the models although PL is slightly better in reproducing the
trend. The dependence of $R_{AA}$ on centrality is better described by PL model. 

{\color{black}
  Figure~\ref{Fig:bJetRAAPt_CMS_276TeV} shows the b jet $R_{AA}$ as a function of jet \ptx in
  several collision centrality bins in PbPb collisions at $\sqrt{s_{NN}}$ = 2.76 TeV
  measured by the CMS experiments~\cite{CMS:2013qak}. The measurements are compared with
  the calculations using power law and GLV energy loss. The heavy quark initiated jets
  are expected to loose less energy than the light quark initiated jets due to the dead
  cone effect~\cite{Dokshitzer:2001zm} however as seen from the Fig.~\ref{Fig:bJetRAAPt_CMS_276TeV}
  the sensitivity in the data is low due to the large statistical uncertainties
  on the measurement~\cite{CMS:2013qak} and the data can be described well with the
  same parameters those are used for inclusive jets in Fig.~\ref{Fig:JetRAAPt_ATLAS_276TeV}
  and Fig.~\ref{Fig:IncJetRAAPt_CMS_276TeV}. The value of quark fraction ($f_{q}$) is taken as
  unity for the GLV calculations in case of b quark R$_{AA}$.
}

%It can be noted that, the $R_{\rm AA}$ requires slightly smaller values of
%$M$ and $\alpha$ as compared to those required for $X_{\rm J}$. It means that the
%broadening in $X_{\rm J}$ still needs larger energy loss over what is needed for $R_{\rm AA}$.

Figure~\ref{Fig:ATLASXJPtBinsCent010at502TeV} shows the 
unfolded Di-jet asymmetry (X$_{\rm J}$) in different $p_{\mathrm{T}}$ bins
in PbPb collisions at $\sqrt{s_{NN}}$ = 5.02 TeV in the most central 0-10$\%$ collisions
measured by ATLAS experiment~\cite{ATLAS:2022zbu} along with the calculations
using power law and GLV energy loss.
The values of the parameters obtained for ATLAS experiment at $\sqrt{s_{NN}}$ = 5.02 TeV
are $\alpha = 0.40$ and $M = 0.32$ GeV/fm for PL and $T_{\rm eff} = 220$ MeV for GLV
which are more as compared to the case for $\sqrt{s_{NN}}$ = 2.76 TeV showing
more energy loss of jets at higher collisions energy. 
Figure~\ref{Fig:ATLASXJPtBinsCent4060at502TeV} shows the 
unfolded Di-jet asymmetry ($X_{\rm J}$) in different $p_{\mathrm{T}}$ bins
in PbPb collisions at $\sqrt{s_{NN}}$ = 5.02 TeV in the peripheral 40-60$\%$ collisions
measured by ATLAS experiment~\cite{ATLAS:2022zbu} along with the calculations
using power law and GLV energy loss.
%The parameters obtained are $\alpha = 0.38$ and $M = 0.46$ GeV/fm for PL
%and $T_{\rm eff} = 330$ MeV for GLV.
The Fig.~\ref{Fig:ATLASXJPtBinsCent010at502TeV} and Fig.~\ref{Fig:ATLASXJPtBinsCent4060at502TeV}
show that PL describe the data much better than the GLV formalism.
Figure~\ref{Fig:JetRAAPt_ATLAS_502TeV} shows the
jet $R_{AA}$ as  a function of jet \ptx in several collision centrality bins
in PbPb collisions at $\sqrt{s_{NN}}$ = 5.02 TeV measured by the ATLAS
experiment~\cite{ATLAS:2018gwx}. The measurements are compared with calculations
using power law and GLV energy loss.
%The parameters obtained are $\alpha = 0.40$ and $M = 0.48$ GeV/fm for PL
%and $T_{\rm eff} = 330$ MeV for GLV.
The dependence of $R_{AA}$ on $p_T$ and centrality is described by both
the models within the experimental errors.

Table~\ref{Tab:ParValues} summarises the values of parameters used to describe 
$A_{J}$ and $X_{J}^{\gamma}$ by CMS experiment at {\SNNO} and 
and $X_{J}$  and $R_{\rm AA}$ by ATLAS experiments at {\SNNO} and {\SNNT}.
In principle, we should need only two sets of values for parameters $\alpha$ and $M$
for two energies. The CMS data of $A_{J}$ and $X_{J}^{\gamma}$ requires larger values of $M$
as compared to ATLAS data. The experimental effects might contribute to the broadening of $X_{\rm J}$
distributions mimicking larger energy loss. 
The ATLAS experiment reportedly removes the effects of experimental resolution and
underlying event to appear in their measured $X_{\rm J}$ spectra. 
%Also $R_{\rm AA}$ requires slightly smaller values of $M$ and $\alpha$ as compared to
%those required for $X_{\rm J}$ for ATLAS experiment. It means that the
%broadening in $X_{\rm J}$ still needs larger energy loss over what is
%needed for $R_{\rm AA}$. Since $R_{\rm AA}$ is the ratio of two quantities, many of
%the experimental effects and systematics are expected to cancel out. 
The extracted values of $\alpha$ are slightly smaller than the $\sqrt{E}$
dependence expected from the BDPS formalism~\cite{Baier:1994bd}.
The parametric analysis from Ref.~\cite{Spousta:2016agr} gives the
value of this index as 0.52. Our analysis show that the energy loss of jets
increases less rapidly with the jet energy than expected from
$\sqrt{E}$ dependence.
The value of $T_{\rm eff}$ is more for higher collision energy which is expected
to create a system at higher temperature. 
The value of $M$ in PL is also dependent on the temperature of the system and
is larger for higher collision energy. 

%Figure~\ref{Fig:dEdxComp_MM32} shows energy loss per unit lenght, dE/dx, as
%a function of jet \ptx in central 0-10$\%$ and peripheral 50-60$\%$ collisions
%for both PL and GLV models.
% The value of dE/dx is similar for both methods in the 0-10$\%$ centrality bin
%while for 50-60$\%$ bin the value for GLV method is smaller becasue of small L$_{\rm eff}$.
%This difference in energy loss dependence of centrality in the two models is
%vissible in figures \ref{Fig:JetRAAPt_ATLAS_276TeV} and \ref{Fig:JetRAAPt_ATLAS_502TeV}.
% The value of dE/dx increases slowly as a function of jet \ptx for both PL and GLV
%models. This gives a slighly increasing trend in $R_{\rm AA}$ as a function of \ptx
%as shown in figures \ref{Fig:JetRAAPt_ATLAS_276TeV} and \ref{Fig:JetRAAPt_ATLAS_502TeV}.
%Overall, \ptx and centrality dependence of  $R_{\rm AA}$ and asymmetry is described better
%by PL model as compared to GLV.

{\color{black}Figure~\ref{Fig:dEdxComp_MM32} shows energy loss per unit lenght, dE/dx, as
a function of jet \ptx in central 0-10$\%$ and peripheral 50-60$\%$ collisions
for both PL and GLV models. The value of dE/dx is more for GLV in the 0-10$\%$
centrality bin while for 50-60$\%$ bin the value for GLV method is smaller becasue of
small L$_{\rm eff}$.}
This difference in energy loss dependence of centrality in the two models is
vissible in figures \ref{Fig:JetRAAPt_ATLAS_276TeV} and \ref{Fig:JetRAAPt_ATLAS_502TeV}.
 The value of dE/dx increases slowly as a function of jet \ptx for both PL and GLV
models. This gives a slighly increasing trend in $R_{\rm AA}$ as a function of \ptx
as shown in figures \ref{Fig:JetRAAPt_ATLAS_276TeV} and \ref{Fig:JetRAAPt_ATLAS_502TeV}.
Overall, \ptx and centrality dependence of  $R_{\rm AA}$ and asymmetry is described better
by PL model as compared to GLV.

%Figure~\ref{Fig:ATLASXJPtBinsPPat502TeV} shows the
%the unfolded Di-jet asymmetry (X$_{\rm J}$) in different $p_{\mathrm{T}}$ bins
%in pp collisions at $\sqrt{s}$ = 5.02 TeV measured by ATLAS experiment~\cite{ATLAS:2022zbu} along
%with the calculations of our model.
%The dependence of $R_{AA}$ on $p_T$ is correctly
%reproduced by both the models. 
% The dependence of $R_{AA}$ on centrality is not so good for the GLV model. 

\section{Summary}
\label{Sec:Summary}

In this work, we quantify the jet energy loss in the medium using a variety of
jet energy loss observables.
The calculations are performed using specific energy loss, $dE/dx$, having
jet energy dependence to be a power law form (PL) or logarithmic form (GLV).
A Monte Carlo model is employed to obtain transverse momentum \ptx and path-lengths
of the initial jets which suffer energy loss in the medium using the above functional forms.
The dijet asymmetry distributions and nuclear modification factors $R_{AA}$ are calculated
and compared with the CMS and ATLAS measurements in Pb+Pb collisions at energies
$\sqrt{s_{\rm NN}}$ = 2.76 TeV and 5.02 TeV.
The dijet asymmetry distributions broaden as the \ptx of the leading jet
increases and/or the collisions are more central, a trend which is reproduced by
Monte Carlo model.
The shapes of asymmetry distributions in all kinematics regions are
better described by power law as compared to GLV form.
The data from CMS requires larger energy loss as compared to ATLAS data to explain the
broadening of the asymmetry ($A_{\rm J}$/$X_{\rm J}$) distributions signifying
the importance of removing experimental effects on the broadening. 
The energy loss increases slightly as one moves from the collision energy 
$\sqrt{s_{\rm NN}}$ = 2.76 TeV to 5.02 TeV as the medium created at higher energy
is supposed to have larger temperature. 
%The energy loss required to explain the nuclear modification factors is found to be slightly
%less as compared to that required to describe $R_{AA}$.
%Since $R_{\rm AA}$ is the ratio of two quantities, many of
%the experimental effects and systematics are expected to cancel out. 
While, the dependence of $R_{AA}$ on \ptx is correctly reproduced by both the power law and
GLV forms, the dependence of $R_{AA}$ on centrality is slightly
better described by PL model. This indicates the dependence of energy loss on
jet energy and path-length is better described by PL model.  
The method can be used to predict the jet $R_{AA}$ and dijet asymmetry values
in heavy ion collisions for future experiments.

%\section*{References}
%\bibliographystyle{iopart-num}
\bibliography{DiJet}

\end{document}